\documentclass[journal]{IEEEtran}
\usepackage[dvips]{graphicx}
\usepackage{url}
\usepackage{cite}
\usepackage{times,amsmath,epsfig}
\usepackage{fancyhdr}
\usepackage{amsmath}
\usepackage{amsfonts}
\usepackage{amssymb}
\usepackage[latin1]{inputenc}
\usepackage{array}
\usepackage{graphicx}
\usepackage{url}
\usepackage{subfigure}
\usepackage{bm}
\usepackage{breqn}
\usepackage{xcolor}
\usepackage{soul}
\usepackage{amssymb}
\usepackage{multirow}

\newcommand{\vct}[1]    {\mbox{\boldmath{$#1$}}}

\newcommand{\dd}        {{\rm d}}

\begin{document}
\title{Stepped-frequency DORT Imaging with Singular Value Decomposition for a Single-antenna Ultrawideband Radar in Multi-path Environments}
\author{Takuya~Sakamoto,~\IEEEmembership{Senior Member,~IEEE,}
  \thanks{This work was supported in part by SECOM Science and Technology Foundation, JSPS 19H02155 and 21H03427, JST JPMJPR1873, and JST COI JPMJCE1307.}
  \thanks{T.~Sakamoto is with the Department of Electrical Engineering, Graduate School of Engineering, Kyoto University, Kyoto, Kyoto, 615-8510 Japan.}}
\markboth{}%
{Sakamoto: Stepped-frequency DORT Imaging for a Single-antenna Ultrawideband Radar in Multi-path Environments}

\maketitle

\begin{abstract}
  Ultrawideband radar is an attractive technology for a variety of applications including security systems. As such, it is essential to develop low-cost systems that produce clear target images. Electromagnetic inverse scattering with time-reversal imaging has been studied for a variety of applications. The time-reversal method, however, uses large-scale antenna arrays, making the system potentially more costly. In this study, we propose an ultrawideband radar imaging algorithm, namely the stepped-frequency DORT (D\'ecomposition de Op\'erateur de Retournement Temporel) algorithm, that uses multi-path scattering for a single antenna. The proposed imaging method is an extension of the conventional DORT method, and uses a frequency-frequency matrix that is suitable for a system with a single antenna. The performance of the proposed method is verified through numerical simulations.
\end{abstract}
\IEEEpeerreviewmaketitle

\section{Introduction}
\IEEEPARstart{U}{ltrawideband} radar is a promising technology for use in a variety of applications including surveillance systems. The clear imaging of surveillance targets requires the development of reliable high-resolution radar imaging systems. In the pursuit of such systems, the time-reversal (TR) UWB radar imaging method has been developed for high-resolution imaging \cite{liu,cresp}. The TR method is based on Lorentz reciprocal theory. The basis of this theory is that by transmitting a time-reversed signal from the receiving antennas, the multiple signals propagating along different paths become coherent at the target location and transmitting antennas, generating a large intensity at these locations at a certain time. In general, this back-propagation process is numerically performed by a computer. It is known that the TR method achieves super-resolution that is much better than the classical resolution limit determined by the antenna aperture. In particular, a clutter-rich environment contributes to better resolution.

To improve the resolution of the TR method, the DORT method was first developed in acoustics and later applied, assuming an antenna array and monochromatic sinusoidal signals, to radar systems by Devaney \cite{devaney}. The DORT method provides a high-resolution capability by separating multiple propagation paths through the application of singular value decomposition (SVD) to a multistatic data matrix generated from bi-static measurements with the array antenna \cite{micolau,fromm,simko}. The columns and rows of this matrix correspond to the transmitting and receiving elements, respectively. The DORT algorithm has also been extended to use wideband signals \cite{yavuz2} and to improve the resolution by introducing a weight term \cite{nguyen}. The multistatic data matrix of the DORT method can be generated in a different manner, by assigning the column and row to the element number and frequency number respectively, as proposed by Yavuz and Teixeira~\cite{teixeira}. Unlike the original DORT method, the modified DORT method can be applied to wideband signals.

These high-resolution DORT methods require antenna array systems, making the system costly and impractical. It is imperative to simplify the system and lower costs if these methods are to be applied to actual security systems; clearly, the number of antennas must be reduced. In radar imaging, Jofre et al.~\cite{jofre} showed that the number of antennas affects the image quality. To remove this lower bound, we use multipath echoes for imaging assuming that the multipath environment is known. This paper extends the original DORT algorithm so that it can be applied to a simplified system with a single antenna in a multipath environment. Moura and Jin \cite{moura} proposed a time-reversal method using a single antenna, but this method was used only for detection and not for imaging. Our method, which we call the stepped-frequency DORT method, generates a matrix to be decomposed by SVD solely in the frequency domain.

Yavuz and Teixeira \cite{yavuz} discussed image deterioration due to the difference between the assumed propagation model and the actual environment. Like the original DORT method, the stepped-frequency DORT method is based on scattering, but from a point-like target instead of a finite-sized target. It is therefore important to establish the performance of the stepped-frequency DORT method for a finite-sized target. We quantitatively evaluate the imaging performance of the method for a target with varying size to investigate the feasibility of the method in practice. Preliminary results of this study have been published in \cite{DORT1,DORT2,DORT3}.

\section{System Model}
\begin{figure}[bt]
  \begin{center}
    \epsfig{file=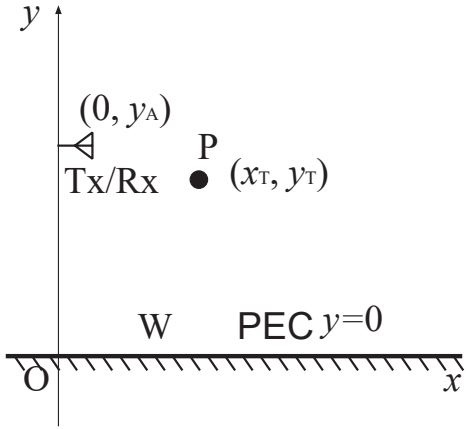,width=0.7\linewidth}    
    \caption{System model of a multipath scattering UWB radar.}
    \label{model}
  \end{center}
\end{figure}

Figure \ref{model} shows the setup of the system assumed in this study; i.e., a two-dimensional system comprising a transverse magnetic wave transmitter/receiver is used to estimate the two-dimensional position of a metallic target. This system comprises a transmit antenna Tx, a receiving antenna Rx, a perfect electric conductor (PEC) plate W, and a point-like target P. The plate W and antenna are on the $x$- and $y$-axes, respectively. In particular, we assume a monostatic radar system with the positions of Tx and Rx in our system the same as those in Fig.~\ref{model}.

In TR imaging and DORT imaging, propagation is expressed with a Green function, which includes multipath effects. The transmitted signal is a UWB pulse $s_{\rm T}(t)$, where $t$ is time. We assume that the distance between the antennas Tx/Rx and the plate W are known. This assumption is not unrealistic as the distance can easily be measured with a strong reflection echo from the wall. The direct wave $s_{\rm D}(t)$ that propagates without scattering from Tx to Rx and the reflected wave $s_{\rm W}(t)$ from plate W are measured and stored in memory prior to actual measurements. Note that $s_{\rm D}(t)$ simply represents a feed point reflection component for monostatic radar because Tx and Rx are located at the same position. These waveforms, $s_{\rm D}(t)$ and $s_{\rm W}(t)$, are subtracted from the received signal $s_0(t)$ to yield $s(t)=s_0(t)-s_{\rm D}(t)-s_{\rm W}(t)$.

The four dominant components in signal $s(t)$ are:
\begin{itemize}
\item $s_1(t)$ Tx-P-Rx,
\item $s_2'(t)$ Tx-P-W-Rx,
\item $s_2''(t)$ Tx-W-P-Rx, and
\item $s_3(t)$ Tx-W-P-W-Rx,
\end{itemize}
where the paths corresponding to $s_2'(t)$ and $s_2''(t)$ are traversed in opposite directions. As a consequence, these echoes cannot be separated. Hereafter, by introducing $s_2(t)=s_2'(t)+s_2''(t)$, only three paths are considered. Note that this model ignores higher-order multiple scattering components.

We next introduce $G(\omega,\vct{r},\vct{r}')$, the Green function associated with propagation through a medium from point $\vct{r}$ to point $\vct{r}'$ including multipath scattering effects expressed as
\begin{equation}
G(\omega,\vct{r},\vct{r}')=G_0(\omega,\vct{r},\vct{r}')+2G_0(\omega,\overline{\vct{r}},\vct{r}')+G_0(\omega,\overline{\vct{r}},\overline{\vct{r}'}),
\end{equation}
where $\omega$ is the angular frequency and $\overline{\vct{r}}$ is the point symmetric with $\vct{r}$ about the $x$-axis, corresponding to a reflection from the metallic plate W or a "mirror image" of the point $\vct{r}$. Moreover, $G_0$ is the Green function of the two-dimensional scalar wave expressed as
\begin{equation}
G_0(\omega,\vct{r},\vct{r}')=\frac{\rm j}{4}H_0\left(\frac{\omega}{c}\left|\vct{r}-\vct{r}'\right|\right),
\end{equation}
where $c$ is the speed of the radio wave, $H_0$ is a Hankel function of the first kind, and $\vct{r}$ and $\vct{r}'$ are the positions of the ends of a propagation path.

Using $S_{\rm T}(\omega)$, the Fourier transform of a transmitted signal $s_{\rm T}(t)$, scattering by a point target can be modeled assuming the Born approximation
\begin{equation}
S(\vct{r}, \omega) = \int \omega^2 C(\vct{r}') G^2(\omega,\vct{r},\vct{r}')S_{\rm T}(\omega)\dd \vct{r}',
\end{equation}
where $\vct{r}$ is an observation point and $C(\vct{r}')$ is a contrast function defined as $C(\vct{r}') = (\varepsilon(\vct{r}')-\varepsilon_0(\vct{r}'))/\varepsilon_0(\vct{r}')$ using the relative permittivity of a target $\varepsilon(\vct{r}')$ and background medium $\varepsilon_0(\vct{r}')$ at position $\vct{r}'$. As explained above, function $G$ includes the multipath propagation effect that enables the high-resolution imaging of the DORT methods.

A monocycle pulse with a central frequency of 4.0 GHz is transmitted, and the received signals are processed for imaging. In particular, the set parameters are: $y_{\rm A}=600.0$ mm, $x_{\rm T}=600.0$ mm and $y_{\rm T}=750.0$ mm, which means that the antenna is at $(0.0 {\rm mm}, 600.0 {\rm mm})$ and the target made of PEC is located at $(600.0 {\rm mm}, 750.0 {\rm mm})$. In this study, the received signals are calculated using the finite difference time domain, with a six-layer perfectly matched layer for absorbing boundaries and a grid resolution of 1.0 mm.

\section{Conventional TR Imaging and DORT Imaging}
\subsection{TR Method}
TR imaging, using the Lorentz reciprocal theorem, is distinguished by its simple signal processing. The principle of TR imaging is described below. $S(\omega)$, the Fourier transform of the received signal $s(t)$ after applying a matched filter $S_{\rm T}^{*}(\omega)$, for a single point target is expressed approximately as
\begin{equation}
S(\omega) = \omega^2 G^2(\omega,\vct{r}_{\rm A},\vct{r}_{\rm P})\left|S_{\rm T}(\omega)\right|^2,
\end{equation}
disregarding constant terms. Here, the positions of Tx and Rx are both taken to be $\vct{r}_{\rm A}$ and the position of the point target is $\vct{r}_{\rm P}$. 

Assume that $s(-t)$ is transmitted from Rx and a strong signal is then received at Tx at $t=0$. This is the basic principle of the TR method. Note that the TR operator $s(t) \rightarrow s(-t)$ is equivalent to complex conjugation in the frequency domain. Therefore, the image $I_{\rm TR}(\vct{x})$ obtained using the TR method is 
\begin{eqnarray}
&I_{\rm TR}&(\vct{x}) = \int \omega^2 S^{*}(\omega)G^2(\omega,\vct{r}_{\rm A},\vct{x})\dd \omega,\\
&=&\!\!\!\!\!\!\int\!\!\omega^4\left|S_{\rm T}(\omega)\right|^2\!{G^*}^2(\omega,\vct{r}_{\rm A},\vct{r}_{\rm P})G^2(\omega,\vct{r}_{\rm A},\vct{x})\dd \omega.
\label{eq:tr}
\end{eqnarray}
$I_{\rm TR}(\vct{x})$ in Eq.~(\ref{eq:tr}) takes its maximum value when $\vct{x}=\vct{r}_{\rm P}$ because the integrand is a real function. As we see here, the TR method is based on matched filter theory.

\subsection{DORT Method}
\begin{figure}[bt]
\begin{center}
\epsfig{file=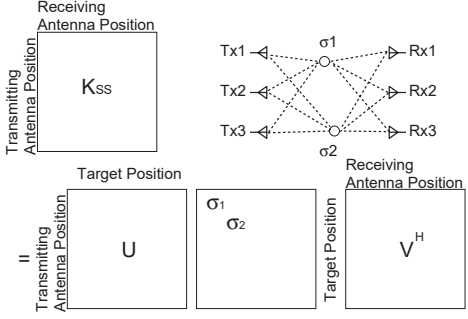,width=0.7\linewidth}
  \caption{SVD of a space--space matrix in the conventional DORT method.}
  \label{matrix}
\end{center}
\end{figure}

DORT imaging is an extension of TR imaging that incorporates SVD to improve the resolution \cite{devaney}. With a space--space matrix $K_{\rm SS}$, called a multistatic data matrix, the DORT method assumes that a sinusoidal wave is transmitted and that there are multiple transmitting and receiving antennas. Element $k_{i,j}$ of $K_{\rm SS}$ is defined as the received complex signal propagating between the $i$-th transmitting antenna and the $j$-th receiving antenna. Here, $k_{i,j}$ is expressed as 
\begin{equation}
k_{i,j}= \sum_{l=1}^{K}\sigma_l g_{i,l} g_{j,l},
\label{kij}
\end{equation}
where $g_{i,l}$ is the Green function for the $i$-th antenna and $l$-th target and $\sigma_l$ is proportional to the scattering intersection of the $l$-th target. The three terms in Eq.~(\ref{kij}) can be divided into three matrices as
\begin{equation}
K_{\rm SS}=U\Sigma V^{\rm H},
\label{svd2}
\end{equation}
where $U$ and $V$ comprise $g_{i,l}$ and $g_{l,j}$, respectively. This decomposition in Eq.~(\ref{svd2}) corresponds to the SVD of $K_{\rm SS}$. Here, $\Sigma$ is a diagonal matrix having diagonal elements $\sigma_l$. The Green function for each propagation path is divided into two matrices $U$ and $V$, thus enabling an imaging similar to the MUSIC method, because we can derive a noise subspace by checking the elements of $\Sigma$. This procedure is illustrated in Fig.~\ref{matrix} for two point-like targets with three transmitting and receiving antennas. Although this method works well in the assumed model with a sinusoidal wave and multiple antennas, it cannot be applied to our system with a single antenna. We therefore introduce the stepped-frequency DORT method in the next section.

\section{Proposed Stepped-frequency DORT}
We assume a single-antenna mono-static radar system as in Fig.~\ref{model}. $S_1,\cdots,S_N$ denotes the respective values of the received signal $S(\omega)$ in the frequency domain at $\omega_1,\omega_2, \cdots,\omega_N$, where $\omega_n = \omega_0 + n\Delta\omega$ for $n = 1, 2,\cdots, N$. The matrix $K_{\rm FF}$ is defined as
\begin{equation}
K_{\rm FF}=
\left[
\begin{array}{cccc}
S_1 & S_2 & \cdots & S_L\\
S_{L+1} & S_{L+2} & \cdots & S_{2L}\\
\vdots & \vdots & \vdots & \vdots \\
S_{N-L+1} & S_{N-L+2} & \cdots & S_{N}\\
\end{array}
\right],
\end{equation}
where the rows and columns respectively correspond to fine and coarse changes in the frequencies: $\omega_i, \omega_{i+1}, \omega_{i+2}, \cdots, \omega_{i+L-1}$ and $\omega_i, \omega_{i+L}, \omega_{i+2L}, \cdots, \omega_{i+N-L}$. The $(i,j)$-th element $k_{i, j} = S_{i+jL+1}$ is defined for $i,j=0, 1,\cdots, L-1$.

We assume $N=L^2$ for simplicity. The Green function can be approximately decomposed into two parts,
\begin{eqnarray}
G_0(\omega_{\rm c}+\omega_{\rm f},\vct{r}',\vct{r})&\simeq& 
-\frac{\rm j}{4}\frac{\exp\left({\rm j}\frac{\omega_{\rm c}+\omega_{\rm f}}{c}\left|\vct{r}-\vct{r}'\right|\right)}
{\sqrt{\frac{\omega_{\rm c}+\omega_{\rm f}}{c}\left|\vct{r}-\vct{r}'\right|}}\nonumber\\
&\simeq& -\frac{\rm j}{4}\frac{\exp\left({\rm j}\frac{\omega_{\rm c}}{c}\left|\vct{r}-\vct{r}'\right|\right)}
{\sqrt{\frac{\omega_{\rm c}}{c}\left|\vct{r}-\vct{r}'\right|}}\nonumber\\
&\cdot&{\exp\left({\rm j}\frac{\omega_{\rm f}}{c}\left|\vct{r}-\vct{r}'\right|\right)},
\end{eqnarray}
where $\omega_{\rm c}$ and $\omega_{\rm f}$ are the coarse and fine frequencies, respectively. 

With this approximation, the Green function for each propagation path can be divided into two parts, namely coarse- and fine-frequency components, with the associated functions forming the basis of the stepped-frequency DORT method. The approximation 
\begin{eqnarray}
k_{i,j} &=& S_{i+jL+1}\nonumber\\
&\simeq& S_{jL+1}\hat{S}_{i}
\end{eqnarray}
thus holds, where $\hat{S}_i$ is the $i$-th adjustment component of the fine-frequency change. Matrix $K_{\rm FF}$ for a single point target is then approximated as 
\begin{eqnarray}
K_{\rm FF}&=&\!\!\!
\left[\!\!\!
\begin{array}{cccc}
S_1 &\!\!\! S_1\hat{S}_1 &\!\!\! \cdots &\!\!\! S_1\hat{S}_{L-1}\\
S_{L+1} &\!\!\! S_{L+1}\hat{S}_1 &\!\!\! \cdots &\!\!\! S_{L+1}\hat{S}_{L-1}\\
\vdots & \vdots & \vdots & \vdots \\
S_{N-L+1} &\!\!\! S_{N-L+1}\hat{S}_1 &\!\!\! \cdots &\!\!\! S_{N-L+1}\hat{S}_{L-1}\\
\end{array}
\right]\nonumber\\
&=&\sigma
\left[
\begin{array}{c}
S_1/\sqrt{\sigma_{\rm L}} \\
S_{L+1}/\sqrt{\sigma_{\rm L}} \\
\vdots \\
S_{N-L+1}/\sqrt{\sigma_{\rm L}} 
\end{array}
\right]
\left[
\begin{array}{c}
\hat{S}_0^{*}/\sqrt{\sigma_{\rm R}}\\
\hat{S}_{1}^{*}/\sqrt{\sigma_{\rm R}}\\
 \vdots\\ 
\hat{S}_{L-1}^{*}/\sqrt{\sigma_{\rm R}} 
\end{array}
\right]^{\rm H}
\label{eq16}\nonumber\\
&=&
\sigma\vct{u}\vct{v}^{\rm H},
\label{suv}
\end{eqnarray}
where the superscript H denotes the Hermitian transpose, $\hat{S}_0=1$ is defined, and $\sigma=\sqrt{\sigma_{\rm L}\sigma_{\rm R}}$ is set to normalize the two vectors $\vct{u}$ and $\vct{v}$ of Eq.~(\ref{suv}). Because this approximation is valid for each target, in considering linear operations, matrix $K_{\rm FF}$ can be decomposed by applying SVD as
\begin{equation}
K_{\rm FF}=U\Sigma V^{\rm H},
\end{equation}
where $\Sigma$ is a diagonal matrix with singular values. The left and right singular matrices $U$ and $V$ correspond to the components for coarse and fine frequencies, respectively.

\begin{figure}[bt]
\begin{center}
\epsfig{file=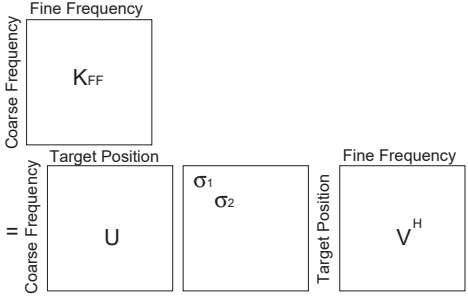,width=0.7\linewidth}
  \caption{SVD of a frequency--frequency matrix in the stepped-frequency DORT method.}
  \label{matrix3}
\end{center}
\end{figure}

As in the conventional DORT method, we adopt small $L-PK$ singular values to estimate noise subspaces, calculating $\vct{u}_{PK+1}\cdots\vct{u}_{N}$ and $\vct{v}_{PK+1}\cdots\vct{v}_{N}$, where $P$ is the number of propagation paths for each point-like target and $K$ the number of targets, as base vectors for a noise subspace. For comparison with the conventional DORT method, Fig.~\ref{matrix3} shows the SVD of $K_{\rm FF}$.

The image from the left singular vectors is
\begin{equation}
\displaystyle
I_{\rm L}(\vct{x}) = \frac{1}{
\displaystyle \sum_{i=PK+1}^{L}\sum_{p=1}^{P}\left|\vct{u}^{\rm H}_i \vct{g}_p(\vct{x})\right|^2/\left|\vct{g}_p(\vct{x})\right|^2},
\label{dortimage}
\end{equation}
where $\vct{g}_p$ is an $L$-dimensional vector constructed from values of the Green function for the $p$-th path at $\omega_1,\omega_{L+1},\omega_{2L+1}\cdots,\omega_{N-L+1}$ written as
\begin{equation}
\vct{g}_p = \left [
\begin{array}{c}
{\omega_1}^2 {G_p}^2({\omega_1},\vct{r},\vct{x})S_{\rm T}({\omega_1})\\
{\omega_{L+1}}^2 {G_p}^2({\omega_{L+1}},\vct{r},\vct{x})S_{\rm T}({\omega_{L+1}})\\
{\omega_{2L+1}}^2 {G_p}^2({\omega_{2L+1}},\vct{r},\vct{x})S_{\rm T}({\omega_{2L+1}})\\
\vdots\\
{\omega_{N-L+1}}^2 {G_p}^2({\omega_{N-L+1}},\vct{r},\vct{x})S_{\rm T}({\omega_{N-L+1}})
\end{array}
\right],
\label{gp}
\end{equation}
with $S_{\rm T}(\omega)$ being the Fourier transform of the transmitted waveform. We define $G_p$ as the Green function for the $p$-th path. In our calculation, a reflection from a metallic wall W is used as an approximation of $\vct{S_{\rm T}}(\omega)$. The waveform is calculated using the finite difference time domain method and stored in advance as reference data. The image $I_{\rm R}(\vct{x})$ can be obtained similarly from the right singular vectors $\vct{v}_{PK+1}\cdots\vct{v}_{N}$ as
\begin{equation}
\displaystyle
I_{\rm R}(\vct{x}) = \frac{1}{
\displaystyle \sum_{i=PK+1}^{L}\sum_{p=1}^{P}\left|\vct{v}^{\rm H}_i \vct{h}_p(\vct{x})\right|^2/\left|\vct{h}_p(\vct{x})\right|^2},
\label{dortimage}
\end{equation}
where $\vct{h}_p$ is an $L$-dimensional vector constructed from values of the Green function for the $p$-th path at $\omega_1,\omega_{2},\omega_{3}\cdots,\omega_{L}$ and written as
\begin{equation}
\vct{h}_p = \left [
\begin{array}{c}
{\omega_1}^2 {G_p}^2({\omega_1},\vct{r},\vct{x})S_{\rm T}({\omega_1})\\
{\omega_{2}}^2 {G_p}^2({\omega_{2}},\vct{r},\vct{x})S_{\rm T}({\omega_{2}})\\
{\omega_{3}}^2 {G_p}^2({\omega_{3}},\vct{r},\vct{x})S_{\rm T}({\omega_{3}})\\
\vdots\\
{\omega_{L}}^2 {G_p}^2({\omega_{L}},\vct{r},\vct{x})S_{\rm T}({\omega_{L}})
\end{array}
\right].
\label{hp}
\end{equation}
We obtain the final image from their product $I_{\rm DORT}(\vct{x})=I_{\rm L}(\vct{x})I_{\rm R}(\vct{x})$.

Note that the existence of a PEC plate has been assumed. However, the proposed method can be used for any multipath environment as long as the Green function of the environment is known. In the following section, the system model with a PEC plate is assumed as the simplest multipath environment for simplicity.

The definition of the matrix $K_{\rm SS}$ in Eq~(9) is not the only way to form a matrix to be decomposed to produce images. There are other ways to form a matrix for our purpose if the matrix can be decomposed into coarse- and fine-frequency components. One possible matrix is found for the forward/backward method \cite{pillai}.

\section{Performance Evaluation of Imaging Methods}
The results of applying the conventional TR and stepped-frequency DORT methods are given in this section. In this paper, we assume $P=3$ and $K=1$ for a simplified model with a single target and a single wall. In the proposed method, we set $L=10$ and $N=100$, while $L-PK=7$ small singular values are selected from the $10\times10$ matrix $K_{\rm FF}$, and the corresponding seven left and right singular vectors are used for imaging. The transmitted pulse has a central frequency of 4.0 GHz and we set $\Delta \omega = 2\pi\times 60$ MHz and $\omega_0=2\pi\times 1.5$ GHz. Note that $\Delta \omega$ should be chosen such that it is not much less than the frequency decorrelation interval because the proposed method cannot generate sufficiently independent singular vectors otherwise. The method of determining the parameter is critical and an important topic of future research.

\begin{figure}[bt]
\begin{center}
\epsfig{file=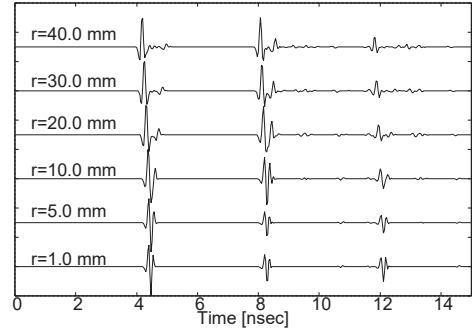,width=0.7\linewidth}
  \caption{Signals received from a target with radius $r$.}
  \label{signals_all}
\end{center}
\end{figure}

For simplicity, the imaging methods are applied to noiseless data. Figure \ref{signals_all} shows the signals received from a cylindrical metallic target with radius $r$. The figure shows three echoes corresponding to the three propagation paths, $s_1(t)$, $s_2(t)$, and $s_3(t)$ defined in Section II. The echoes are received earlier as the radius increases, while creeping echoes are also observed for a target with a larger radius. Note that the waveform distortions generated by the larger targets can degrade the estimated images because the stepped-frequency DORT algorithm assumes a Green function based on Rayleigh scattering at a point target.

\begin{figure}[bt]
\begin{center}
  \epsfig{file=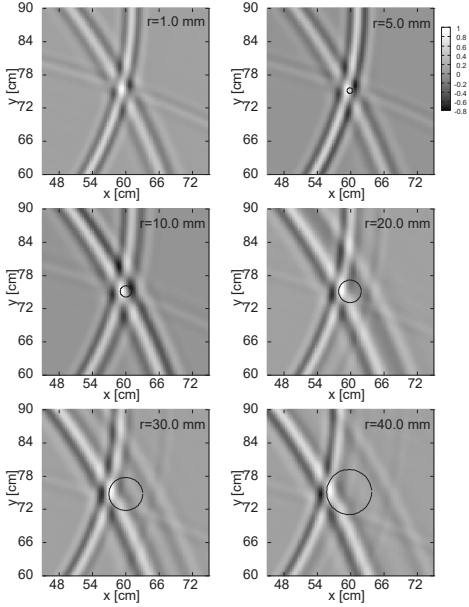,width=0.7\linewidth}
  \caption{Images produced in conventional TR imaging.}
  \label{tr}
\end{center}
\end{figure}

Figure \ref{tr} shows the image estimated using the conventional TR method. In the case of small $r$, three waveforms interfere to generate a prominent peak at the correct position. However, the three waveforms do not meet at the same point for large $r$, and the maximum peak is shifted to the point where two of the waveforms intersect. In addition, we see artifacts generated by creeping waves for large $r$. The estimated target position is close to the target boundary for large $r$.

\begin{figure}[bt]
\begin{center}
\epsfig{file=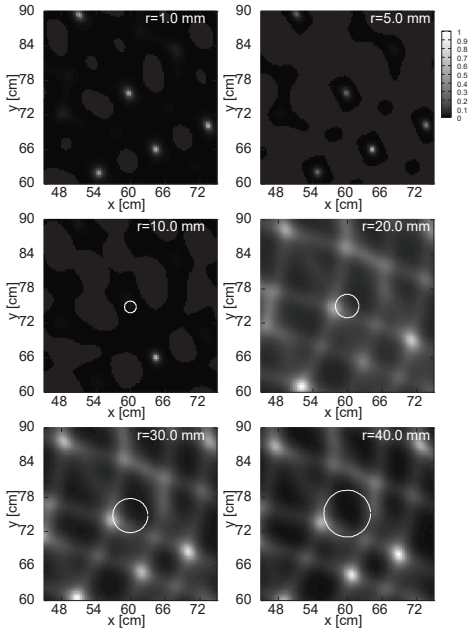,width=0.7\linewidth}
  \caption{Images produced by the left singular vectors.}
  \label{allfine}
\end{center}
\end{figure}

The image from the left singular matrix $I_{\rm L}(\vct{x})$ is shown in Fig.~\ref{allfine}. Although the actual target position is estimated with great sharpness for $r=1.0$ mm and $r=5.0$ mm, there are undesired peaks in the images. For $r=10.0$ mm, only a small response is observed at the actual target position, whereas a strong artifact is seen. For $r\geq 20.0$ mm, the intensity at the actual target location becomes large again, but with compromised sharpness.

The artifacts in these images are interpreted using an analogy from conventional grating lobes in an antenna array pattern with large antenna intervals. The frequency interval of the left singular vectors is $\omega_{L+1}-\omega_{1}$, which is $L$ times the original frequency interval as seen in Eq.~(12). The frequency interval determines the observation time, and multiple periodical artifacts are seen outside of this observation time. Meanwhile, the right singular vectors have the same frequency interval $\omega_{1}-\omega_{0}$ as in Eq.~(12) as the original signal in the frequency domain and they do not produce artifacts unlike the left singular vectors. However, the frequency bandwidth is $S_{L-1}-S_{0}$, which is narrower than that of the original signal by the factor $L$, which worsens the resolution. For these reasons, we see the differences between Figs.~6 and 7.

\begin{figure}[bt]
\begin{center}
\epsfig{file=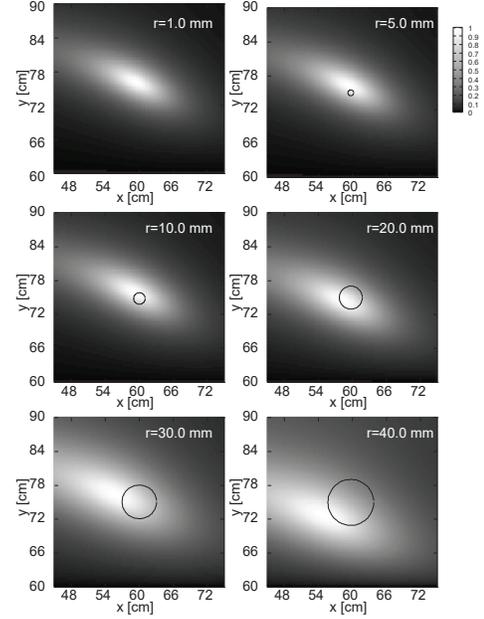,width=0.7\linewidth}
  \caption{Images produced by the right singular vectors.}
  \label{allcoarse}
\end{center}
\end{figure}

The image from the right singular matrix $I_{\rm R}(\vct{x})$ is shown in Fig.~\ref{allcoarse}. Although there are no false images, the resolution is worse than that of $I_{\rm L}(\vct{x})$. In this case, although there are no artifacts, the image quality is also worse than that of $I_{\rm L}(\vct{x})$. Again, for large $r$, the image sharpness deteriorates, with the image quality particularly degraded for $r\geq 30.0$ mm.

\begin{figure}[bt]
\begin{center}
\epsfig{file=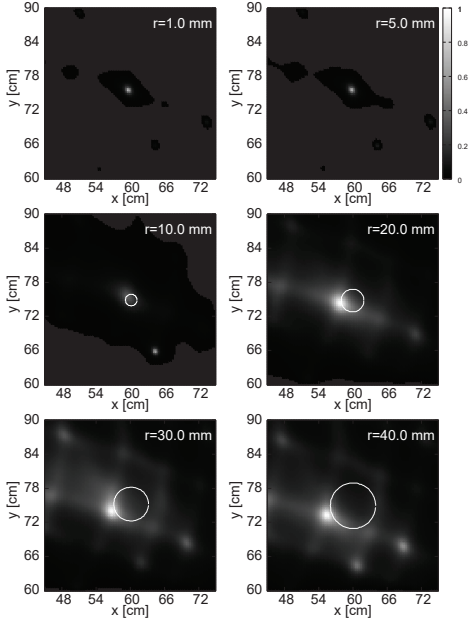,width=0.7\linewidth}
  \caption{Images obtained using the stepped-frequency DORT method.}
  \label{all}
\end{center}
\end{figure}

The final image of the stepped-frequency DORT method is obtained as the product of the two kinds of images $I_{\rm DORT}(\vct{x})=I_{\rm L}(\vct{x})I_{\rm R}(\vct{x})$ to improve the sharpness while suppressing artifacts. Figure \ref{all} shows the images obtained using the stepped-frequency DORT method. The sharpness of $I_{\rm L}(\vct{x})$ and the artifact suppression effect of $I_{\rm R}(\vct{x})$ are incorporated to produce clear images. We see that the images are clear when $r$ is small, although the images are blurred and residual artifacts are produced when $r$ is large. However, the stepped-frequency DORT method is still able to produce an image even for a relatively large target.

\section{Evaluation of Accuracy and Sharpness}
\begin{figure}[bt]
  \begin{center}
\epsfig{file=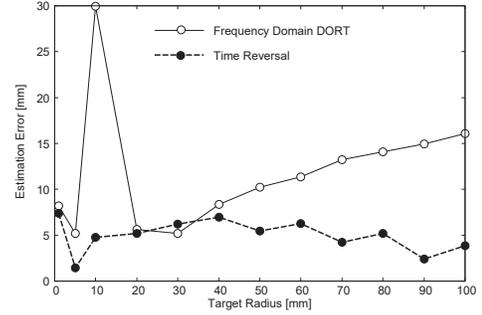,width=0.7\linewidth}
  \caption{Estimation error for each method.}
  \label{error}
  \end{center}
\end{figure}

This section quantitatively investigates the image quality of the methods. Figure \ref{error} shows the estimation error of a target position for the conventional TR method and stepped-frequency DORT method. The error $e(r)$ for a target with radius $r$ is defined as the distance between the estimated position and the point on the target surface that is closest to it, and it is expressed as
\begin{equation}
e(r) = \left|\left|\vct{x}_{\rm E}-\vct{x}_{\rm T}\right|-r\right|
\end{equation}
for the estimated target position $\vct{x}_{\rm E}$ and the actual target center position $\vct{x}_{\rm T}$. Here, the number of targets $K=1$ is assumed to be known. The figure shows that the conventional TR method has an error less than 10.0 mm for any $r$ whereas the stepped-frequency DORT method has a larger error for large $r$. This can be explained in that the orthogonality between the noise subspace and Green function assumed in the stepped-frequency DORT method is not satisfied because the point target model is not valid for actual scattering with a finite-sized target. In addition, the waveform distortion including creeping waves contributes to the degradation of the image. Note that the stepped-frequency DORT method has large error for $r=10.0$ mm because the maximum point erroneously falls on the false artifact as shown in  Fig.~\ref{all}. Such unstable behavior is also seen for other high-resolution techniques, such as the MUSIC and Capon methods.

\begin{figure}[bt]
  \begin{center}
\epsfig{file=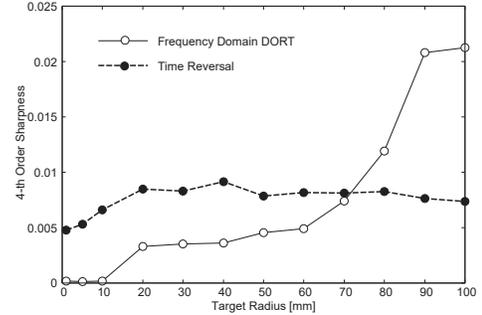,width=0.7\linewidth}
  \caption{Sharpness metric for each method.}
  \label{sharpness}
  \end{center}
\end{figure}

We next evaluate the sharpness of the images using the Muller and Buffington (MB) sharpness metric \cite{mbs}. The $q$-th order of this metric $h_q$ is defined as
\begin{equation}
h_q = \frac{1}{M}\sum_{m=1}^{M} I_m^q,
\end{equation}
where $I_m$ is a vector with elements corresponding to pixel intensities normalized by the maximum pixel intensity of the image, and $M$ is the number of pixels in the image. The exponent $q$ determines the order of the statistics, which is the sharpness of the image for $q>2$ with higher-order statistics. Note that for this metric, small values of $h_q$ signify sharper images. Here, we set $q=4$ and evaluate the sharpness of the images. Figure \ref{sharpness} shows the fourth-order MB sharpness metric for each method. The figure shows that the stepped-frequency DORT method provides high sharpness for small $r$, in particular for $r\leq 10.0$ mm. The stepped-frequency DORT method provides greater sharpness than the conventional TR method for $r\leq 70.0$ mm. Conversely, for $r> 70.0$ mm, the conventional TR method provides greater sharpness. The conventional TR method provides almost constant sharpness regardless of $r$.

Note that the number of targets has been assumed to be known. The performance of the proposed method when the assumed number of targets is not correct can be inferred from the characteristics of the MUSIC method. If the assumed number of targets is larger than the actual number of targets, the proposed method generates false images. If the assumed number of targets is smaller than the actual number, some of the targets cannot be imaged. It is an important future task to investigate the detailed characteristics of the proposed method under these conditions.

\section{Conclusion}
A new method, namely stepped-frequency DORT, was proposed for electromagnetic inverse scattering assuming a UWB radar system. This method is applicable to a wideband radar system with a single antenna. The proposed method was derived by incorporating a frequency--frequency matrix into the conventional DORT algorithm, allowing the method to be applied to measurement with a single antenna, whereas the conventional DORT algorithm assumes a system with multiple antennas or sinusoidal signals. The performance of the proposed method was investigated in numerical simulations. The results show that the proposed method has higher resolution than the conventional TR method.

\section*{Acknowledgment}
We thank Prof. Toru Sato of Kyoto University for his help with this study.

\begin{IEEEbiography}[{\includegraphics[width=1in,height=1.25in,clip,keepaspectratio]{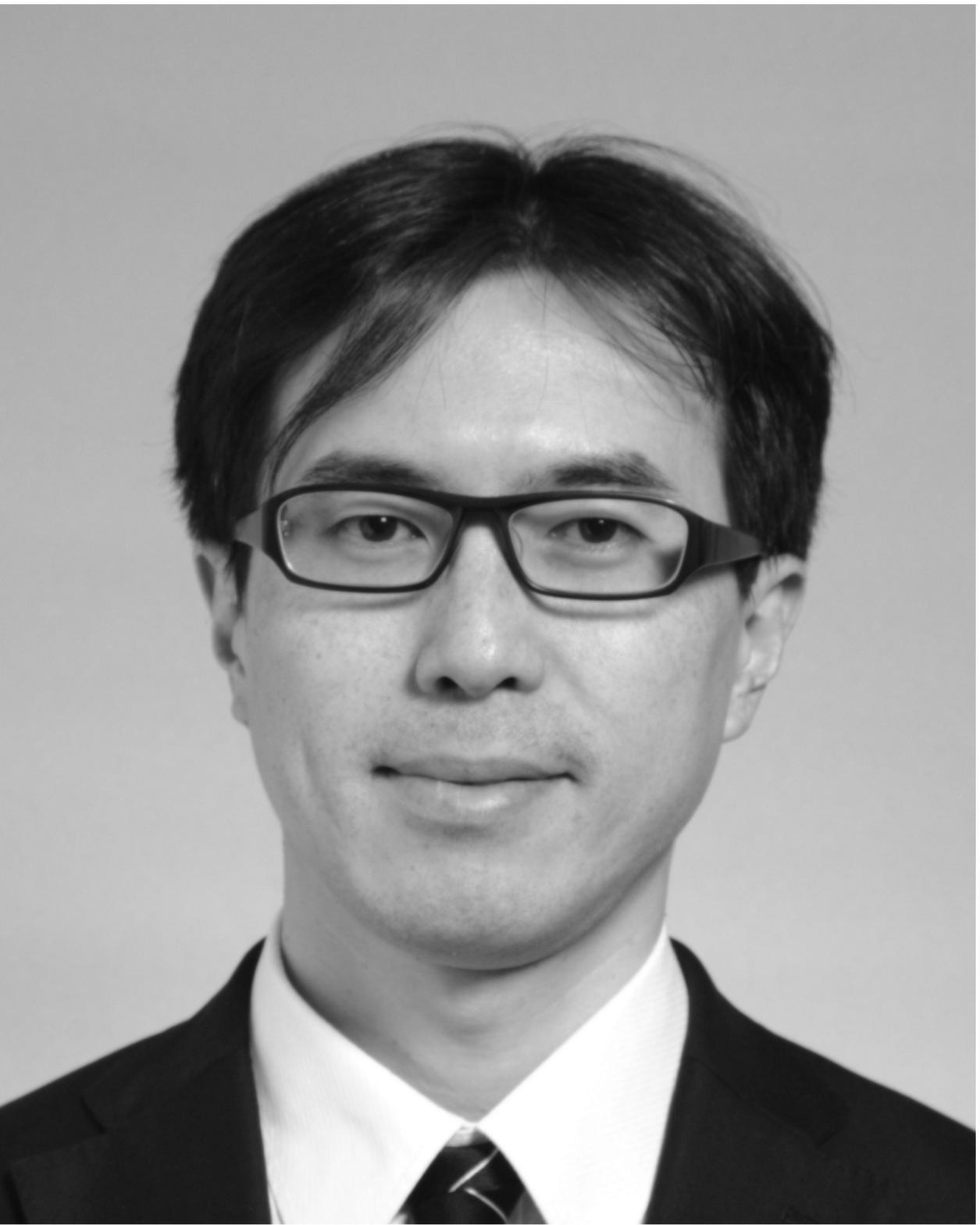}}]{Takuya Sakamoto} (M'04--SM'17) received a B.E.~degree in electrical and electronic engineering from Kyoto University, Kyoto, Japan, in 2000, and M.I.~and Ph.D.~degrees in communications and computer engineering from the Graduate School of Informatics, Kyoto University, in 2002 and 2005, respectively.

  From 2006 to 2015, he was an Assistant Professor at the Graduate School of Informatics, Kyoto University. From 2011 to 2013, he was also a Visiting Researcher at Delft University of Technology, Delft, The Netherlands. From 2015 to 2018, he was an Associate Professor at the Graduate School of Engineering, University of Hyogo, Himeji, Japan. In 2017, he was also a Visiting Scholar at the University of Hawaii at Manoa, Honolulu, HI, USA. Since 2018, he has been a PRESTO Researcher at the Japan Science and Technology Agency, Kawaguchi, Japan. At present, he is an Associate Professor at the Graduate School of Engineering, Kyoto University. His current research interests are system theory, inverse problems, radar signal processing, radar imaging, and wireless sensing of vital signs.

Dr. Sakamoto was a recipient of the Best Paper Award from the International Symposium on Antennas and Propagation (ISAP) in 2012 and the Masao Horiba Award in 2016. In 2017, he was invited as a semi-plenary speaker to the European Conference on Antennas and Propagation (EuCAP) in Paris, France.
\end{IEEEbiography}

\end{document}